\documentclass[sn-basic]{sn-jnl}

\usepackage{graphicx}%
\usepackage{multirow}%
\usepackage{amsmath,amssymb,amsfonts}%
\usepackage{amsthm}%
\usepackage{mathrsfs}%
\usepackage[title]{appendix}%
\usepackage{xcolor}%
\usepackage{textcomp}%
\usepackage{manyfoot}%
\usepackage{booktabs}%
\usepackage{subcaption}%
\usepackage{algorithm}%
\usepackage{algorithmicx}%
\usepackage{algpseudocode}%
\usepackage{listings}%

\theoremstyle{thmstyleone}%
\theoremstyle{thmstyletwo}%

\theoremstyle{thmstylethree}%

\begin{document}
\title{Spontaneous Spatial Cognition Emerges during Egocentric Video Viewing through Non-invasive BCI }

\author[1]{\fnm{Weichen} \sur{Dai}}

\author[1]{\fnm{Yuxuan} \sur{Huang}}

\author[1]{\fnm{Li} \sur{Zhu}}

\author[1]{\fnm{Dongjun} \sur{Liu}}

\author[2]{\fnm{Yu} \sur{Zhang}}

\author[3]{\fnm{Qibin} \sur{Zhao}}

\author[3,4,5]{\fnm{Andrzej} \sur{Cichocki}}

\author[6]{\fnm{Fabio} \sur{Babiloni}}

\author[1]{\fnm{Ke} \sur{Li}}

\author[1]{\fnm{Jianyu} \sur{Qiu}}

\author[1]{\fnm{Gangyong} \sur{Jia}}

\author*[1]{\fnm{Wanzeng} \sur{Kong}}

\author*[1]{\fnm{Qing} \sur{Wu}}

\affil*[1]{\orgname{Hangzhou Dianzi University}, \country{China}}

\affil[2]{\orgname{Zhejiang University}, \country{China}}

\affil[3]{\orgname{RIKEN Center for Advanced Intelligence Project}, \orgname{RIKEN}, \country{Japan}}

\affil[4]{\orgname{Systems Research Institute}, \orgname{Polish Academy of Sciences}, \country{Poland}}

\affil[5]{\orgname{Nicolaus Copernicus University}, \country{Poland}}

\affil[6]{\orgname{University of Rome Sapienza}, \country{Italy}}

\abstract{Humans possess a remarkable capacity for spatial cognition, allowing for self-localization even in novel or unfamiliar environments. While hippocampal neurons encoding position and orientation are well documented, the large-scale neural dynamics supporting spatial representation—particularly during naturalistic, passive experience—remain poorly understood.
Here, we demonstrate for the first time that non-invasive brain–computer interfaces (BCIs) based on electroencephalography (EEG) can decode spontaneous, fine-grained egocentric 6D pose—comprising three-dimensional position and orientation—during passive viewing of egocentric video. Despite EEG’s limited spatial resolution and high signal noise, we find that spatially coherent visual input (i.e., continuous and structured motion) reliably evokes decodable spatial representations, aligning with participants’ subjective sense of spatial engagement. Decoding performance further improves when visual input is presented at a frame rate of 100 ms per image, suggesting alignment with intrinsic neural temporal dynamics.
Using gradient-based backpropagation through a neural decoding model, we identify distinct EEG channels contributing to position- and orientation-specific components, revealing a distributed yet complementary neural encoding scheme.
These findings indicate that the brain’s spatial systems operate spontaneously and continuously, even under passive conditions, challenging traditional distinctions between active and passive spatial cognition. Our results offer a non-invasive window into the automatic construction of egocentric spatial maps and advance our understanding of how the human mind transforms everyday sensory experience into structured internal representations.
}

\maketitle

\section{Introduction}\label{sec1}

\label{sec:intro}

Humans and animals exhibit remarkable spatial cognition through vision, demonstrating a strong ability to sense and localize their positions in unknown environments~\cite{dicarlo2012does,killian2012map,finnie2021spatiotemporal,dotson2021nonlocal}. 
Through invasive techniques, the mammalian hippocampus~\cite{bohbot2017low,goyal2020functionally} and associated brain regions have identified specialized neurons~\cite{alexander2023rethinking,miller2018lateralized}, including spatial view cells~\cite{georges1999spatial,rolls2005spatial}, place cells~\cite{o1971hippocampus,bats2008representation}, head direction cells~\cite{taube1990head,finkelstein2015three}, grid cells~\cite{hafting2005microstructure,ginosar2021locally,wagner2023entorhinal}, and time cells~\cite{tsao2018integrating,issa2020navigating,omer2023contextual}, that related to spatial cognition. These cells provide a spatiotemporal representation of the environment, forming a continuous stream where each moment encodes information about the past, present, and future~\cite{dotson2021nonlocal}. This enables precise and robust navigation, allowing individuals to know where they are, where they want to go, and how to reach their destination\cite{shao2024non}. 

In addition to the level of neurons, the study of spatial cognition mechanisms by analyzing entire brain activities is a common approach~\cite{bonner2017coding,delaux2021mobile}.
Compared to Electrocorticography (ECoG) and functional magnetic resonance imaging (fMRI)~\cite{taube2013navigation,duarte2016anterior,quan2024psychometry}, EEG stands out as a non-invasive technique that is simple, affordable, portable, and user-friendly for collecting physiological electrical signals~\cite{casson2018electroencephalogram}. 
These signals reflect neuronal activity in the cerebral cortex and overall brain function, making EEG a subject of great interest. EEG is widely used in various applications, including emotion recognition~\cite{liu2023brain,alarcao2017emotions}, brain to image~\cite{kavasidis2017brain2image}, event-related potentials~\cite{dietrich2010review}, and motor imagery tasks~\cite{pfurtscheller2006mu,ding2025eeg}.

In spatial behavior research, EEG has emerged as a complementary tool for neuroimaging~\cite{baker2009way,plank2010human,lin2009eeg}, especially under naturalistic conditions, such as indoor and outdoor walking~\cite{ladouce2017understanding,reiser2019recording,maoz2023dynamic}. 
Most studies focus on event-related spectral perturbation (ERSP), particularly in the alpha and theta frequency bands, which are the most extensively studied oscillations~\cite{plank2010human,delaux2021mobile}. These bands have consistently been shown to correlate with mental states, strategies, and stimulus characteristics. Researchers primarily rely on energy analysis to validate the relationship between specific brain regions and spatial cognition, demonstrating that the retrosplenial cortex (RSC) plays a crucial role in translating between egocentric and allocentric spatial information~\cite{gramann2010human,lin2015eeg,long2025allocentric}. 

Despite recent progress, several critical questions continue to hinder a deeper understanding of spatial cognition using EEG-based approaches~\cite{chrastil2022theta,vavrevcka2012frames,delaux2021mobile}.
(1) It remains unclear whether spontaneous spatial representations are reflected in EEG during natural viewing conditions, as most existing studies rely on artificially designed stimuli that may fail to capture natural spatial processing.
(2) It is still debated whether fine-grained 6D pose information is represented in scalp EEG signals, as current approaches often yield only coarse or superficial insights.
(3) The temporal dynamics of spatial cognition in EEG are poorly understood—particularly, what temporal resolution of visual input is optimal for revealing clear spatial representations.
(4) Little is known about the specific contributions of different EEG channels to position- versus orientation-related processing during self-localization, as most studies are constrained by predefined features or frequency-band EEG analyses.

To address these questions, we propose a data-driven approach that directly decodes fine-grained spatial representations from non-invasive EEG signals evoked during passive viewing of egocentric videos with structured visual continuity. Rather than relying on artificial geometric shapes or synthetic visual stimuli, our paradigm leverages naturalistic visual input to engage spontaneous, subconscious spatial cognition. By decoding full 6D pose transformations, we aim to test whether such fine-grained latent spatial information is indeed embedded in EEG signals. The decoding of the 6D pose is conceptually similar to the localization of the pose in robotics, which involves estimating the position and orientation of a camera from one or more images, an allocentric representation of space~\cite{vavrevcka2012frames,bats2008representation}. In our context, it involves regressing 6D poses from EEG signals while subjects view first-person videos as if situated within the scene. 

We further investigate the optimal temporal resolution of visual input for eliciting spatially informative EEG patterns, revealing how the brain encodes dynamic spatial context over time. To uncover the cortical basis of these representations, we apply gradient-based backpropagation to identify EEG channels contributing most to position- and orientation-specific decoding. 

Together, our study bridges low-SNR EEG recordings with fine-grained spatial representation and establishes a non-invasive framework for probing spontaneous spatial processing in naturalistic settings. Our findings demonstrate that EEG signals encode spatial cognition information, even in passive viewing scenarios. This suggests that individuals, even while seated and stationary, subconsciously engage in spatial reasoning and 3D localization in response to visual stimuli. This phenomenon aligns with common experiences, such as feeling disoriented when playing immersive 3D video games or using a smartphone in a moving vehicle, underscoring the automatic and continuous nature of spatial cognition in everyday contexts.

\section{Results}\label{sec2}
\begin{figure}[t!]
    \centering
    \includegraphics[width=0.6\linewidth]{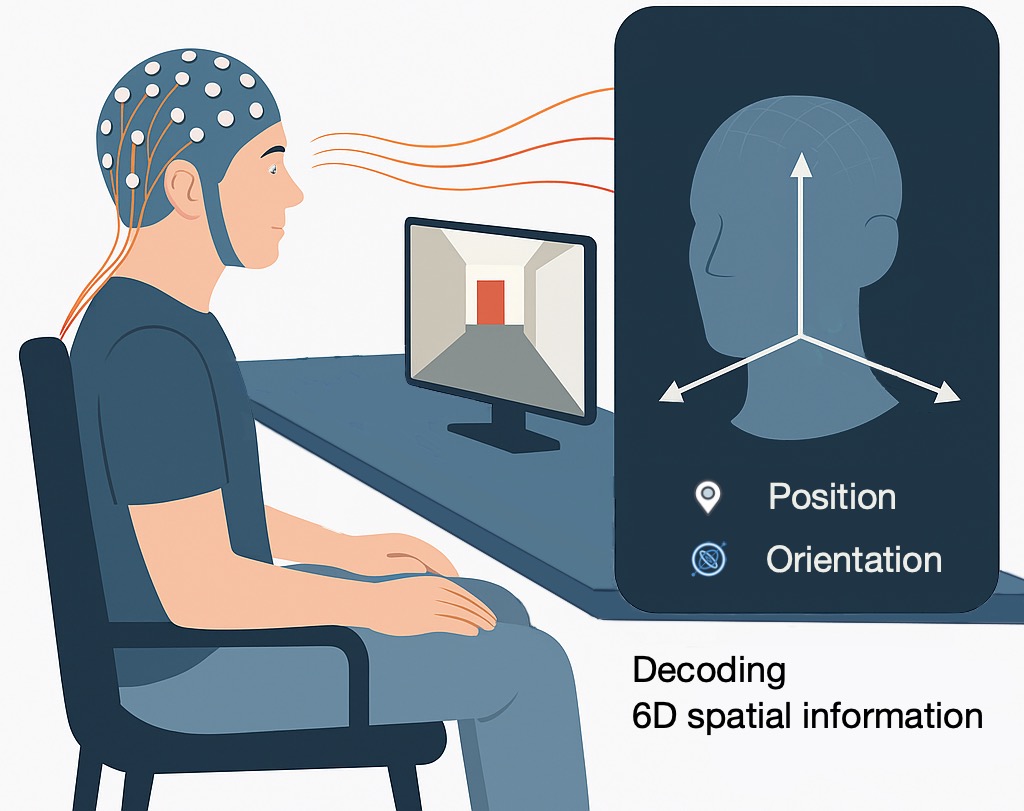}
    \caption{Subjects view first-person perspective video frames simulating immersion within the scene.}
    \label{fig:meth}
\end{figure}

\subsection{Do Spontaneous Spatial Representations Emerge in EEG during Egocentric Video Viewing?}

\begin{figure}[t!]
    \centering
    \includegraphics[width=\linewidth]{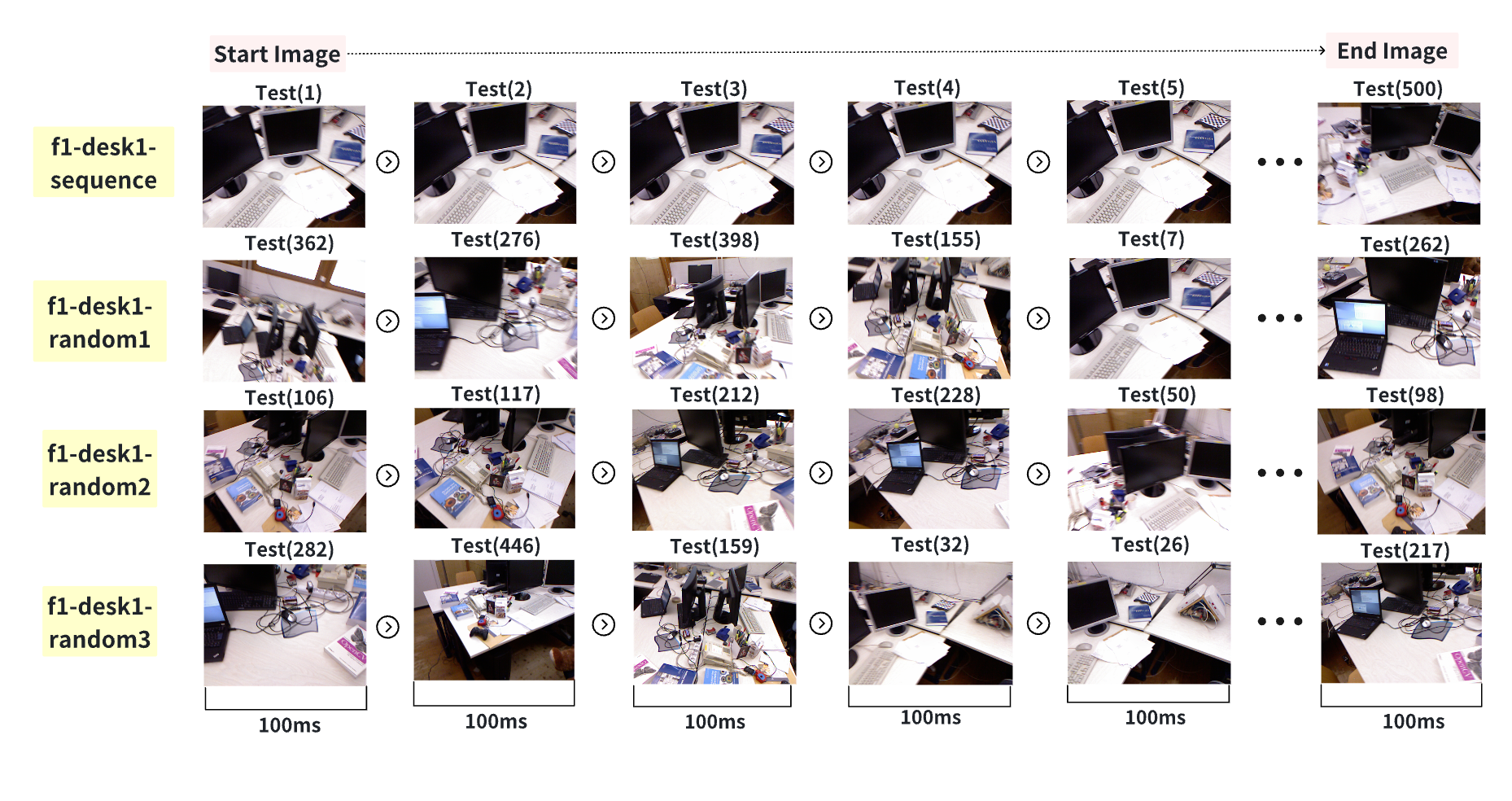}
\caption{Illustration of the image presentation paradigm. 
        Each image in the sequence is presented for 100~ms in default.}
    \label{fig:exp1}
\end{figure}

To investigate whether naturalistic egocentric visual input can elicit spontaneous spatial cognition, and whether such information is embedded in scalp EEG signals, we designed a controlled experiment with two video playback conditions. In the sequential condition, participants viewed egocentric videos with frames presented in their original, chronologically ordered sequence (\textit{sequence}). In the random condition, the same frames were temporally shuffled and shown in a non-chronological order (\textit{random}). An illustration of these playback paradigms is provided in Fig.~\ref{fig:exp1}.
\begin{figure}[!t]
   \centering
\includegraphics[width=\linewidth]{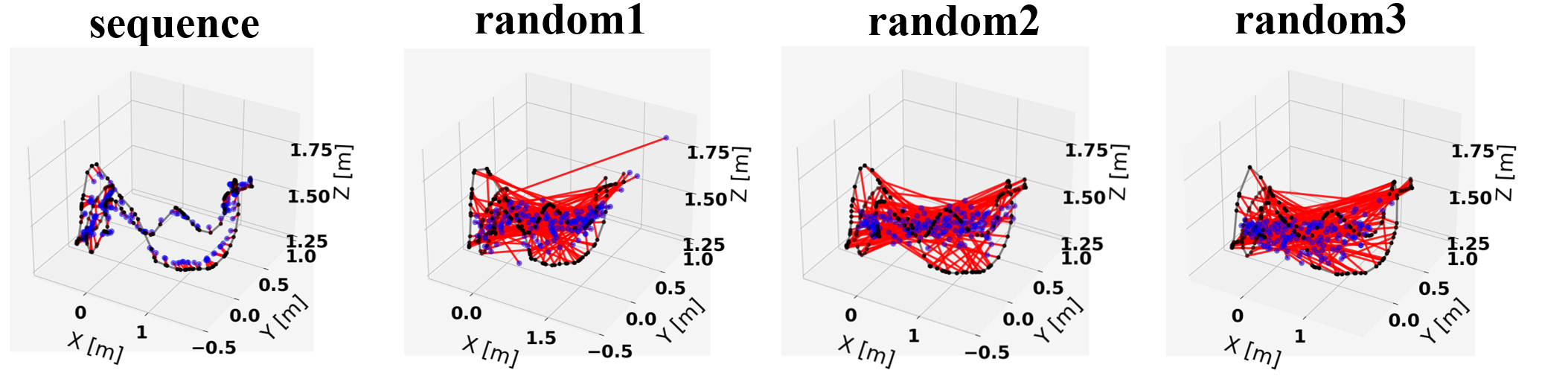}
\caption{Visualization of trajectories under sequential and random playback conditions. The ground truth trajectories are shown as black lines, while the decoded poses are represented by blue lines. The red dashed lines indicate the corresponding errors.}
    \label{fig:ramdom-vis}
\end{figure}

\begin{figure}[htbp]
    \centering
    \includegraphics[width=\linewidth]{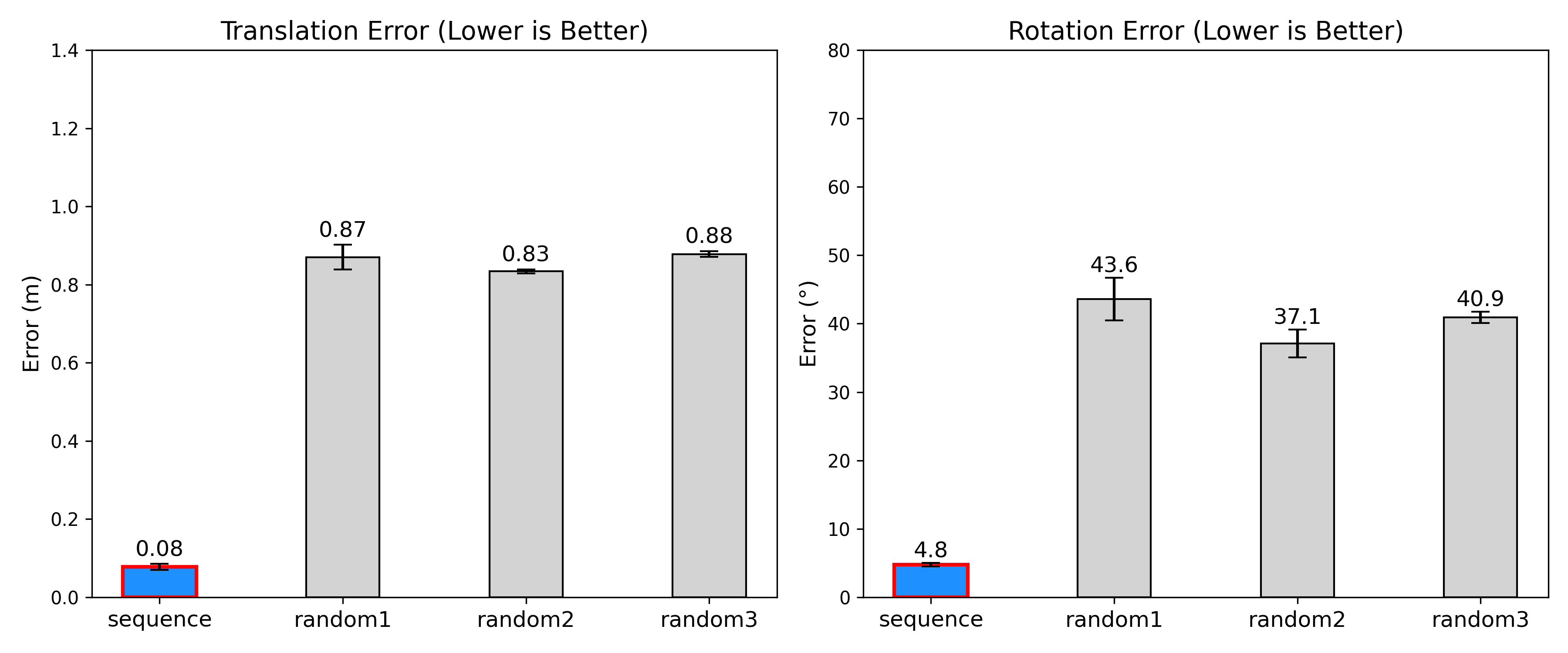}  
    \caption{
    The left panel shows the mean translational errors (in meters) with standard deviation error bars, 
    and the right panel shows the mean rotational errors (in degrees) with corresponding standard deviations. 
    The \textbf{sequence} condition is highlighted with a red border to emphasize its significantly lower errors, 
    indicating that sequential visual stimuli improve the EEG-based pose decoding accuracy and stability compared to random stimuli.
    }
    \label{fig:exp1-ana}
\end{figure}

As shown in Fig.~\ref{fig:exp1-ana}, decoding accuracy of 6D spatial pose from EEG signals was significantly higher in the sequential condition compared to the random condition. Although the decoding framework still produced outputs under the random condition, Fig.~\ref{fig:ramdom-vis} reveals that both our proposed method and standard baselines failed to yield meaningful predictions—errors approached the full range of the trajectory, indicating near-random performance.

These results suggest that temporally coherent visual input plays a critical role in enabling the brain to construct internal spatial representations. When visual frames follow a natural sequence, the brain can integrate temporal and motion continuity cues to infer spatial layout. Conversely, when frames are disordered, this integration is disrupted, impairing spatial reasoning and degrading EEG-based decoding performance.

Notably, participants’ subjective reports echoed the decoding results. Under the random condition, they reported being disoriented and unable to infer the camera's viewing direction. In contrast, sequential playback enabled them to perceive a coherent, immersive spatial environment.

The stark contrast in decoding accuracy between sequential and random conditions indicates that the EEG signals do encode meaningful information related to spatial cognition. This argues against the possibility of spurious decoding results driven solely by machine learning overfitting.

\subsection{Is Fine-Grained 6D Pose Information Represented in Scalp EEG Signals?}
To further investigate whether scalp EEG signals contain fine-grained spatial representations, we conducted experiments in which participants viewed egocentric videos recorded in both indoor and outdoor environments. EEG data were used to decode 6D poses (3D position + 3D orientation).

\begin{figure}[!ht]
    \centering
\includegraphics[width=\linewidth]{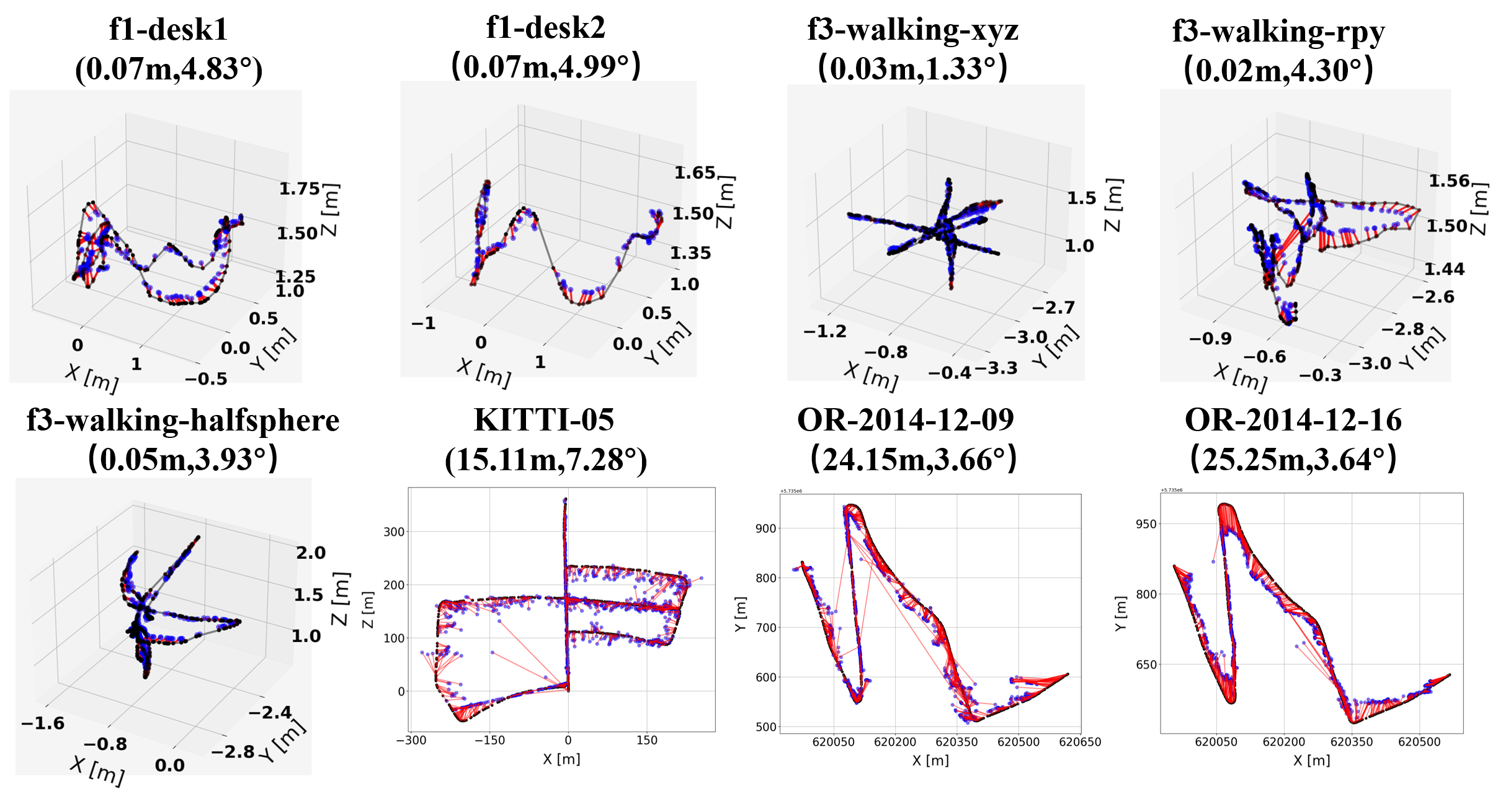}
    \caption{The proposed method on indoor and outdoor datasets. The ground truth trajectories are shown as black lines, while the decoded poses are represented by blue lines. The red dashed lines indicate the corresponding errors. Indoor datasets are visualized in 3D space, while outdoor datasets are visualized from a top-down view. There are average errors (translational error in meters, rotational error in degrees) below the sequence name. }
    \label{fig:combined}
\end{figure}

As shown in Fig.~\ref{fig:combined}, our proposed method approaches—successfully decoded 6D pose information from EEG signals when videos were played in a temporally coherent (i.e., sequential) manner, regardless of whether the environment was indoor or outdoor. These results provide compelling evidence that fine-grained spatial representations, including both translation and rotation, are indeed embedded in scalp EEG signals.

\subsection{How Does the Temporal Resolution of Egocentric Video Impact EEG-Based Spatial Decoding?}
Low frame rates (the speed which images are presented) may cause the loss of information between key frames, introduce temporal gaps and discontinuities, and impair the brain's ability to form a coherent spatial representation, thereby reducing the accuracy of spatial cognition decoding. To evaluate the impact of frame rate on brain spatial cognition, we conducted a series of experiments using different frame rate sequences from the freiburg1-desk1 image sequence in the TUM-RGBD dataset.
\begin{figure}[htbp]
    \centering
    \includegraphics[width=\linewidth]{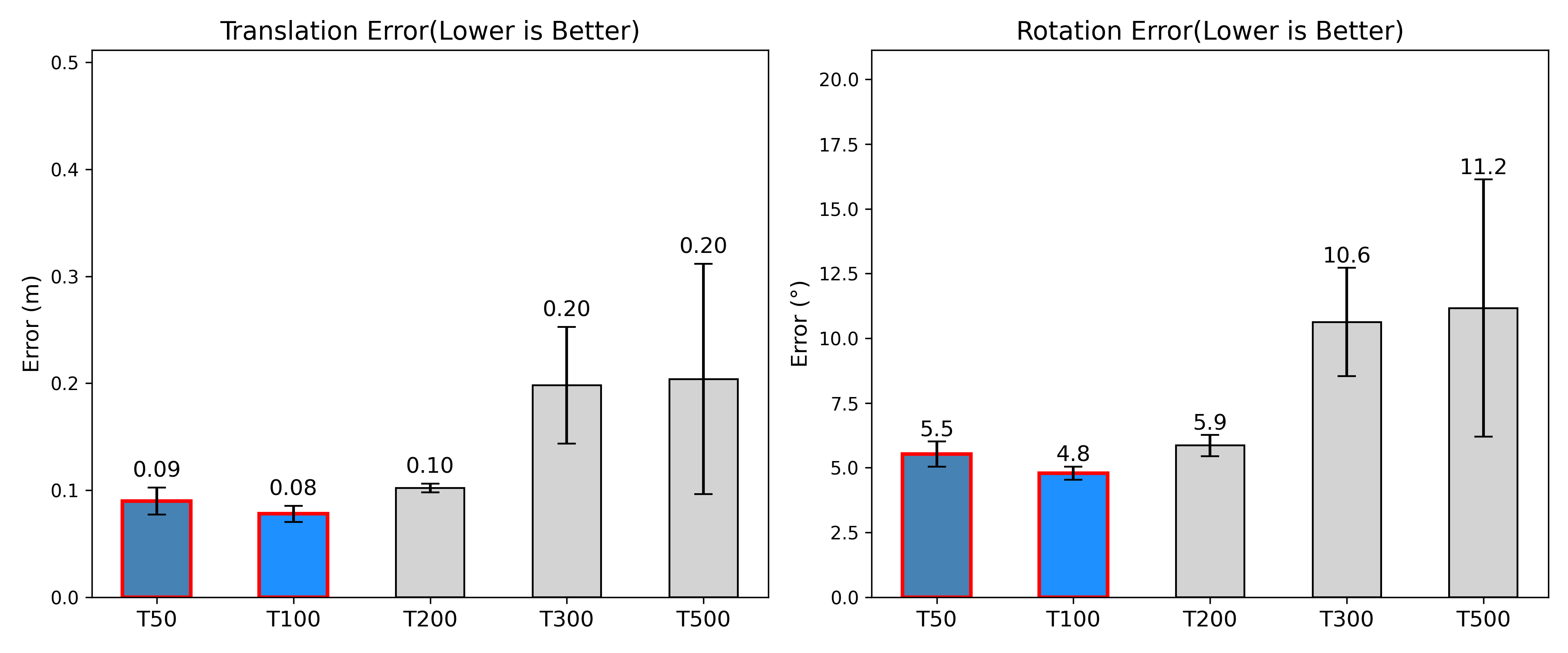}
    \caption{
        Translation error(left) and rotation error(right) at different frame rates(50ms--500ms). Each bar shows the mean error with error bars indicating the standard deviation. The frame rates(50ms and 100ms) are highlighted with red borders to emphasize their superior performance, with 100ms achieving the best decoding accuracy and 50ms following as the second best.
    }
    \label{fig:exp2-analysis}
\end{figure}
\begin{figure}[!ht]
   \centering
\includegraphics[width=\linewidth]{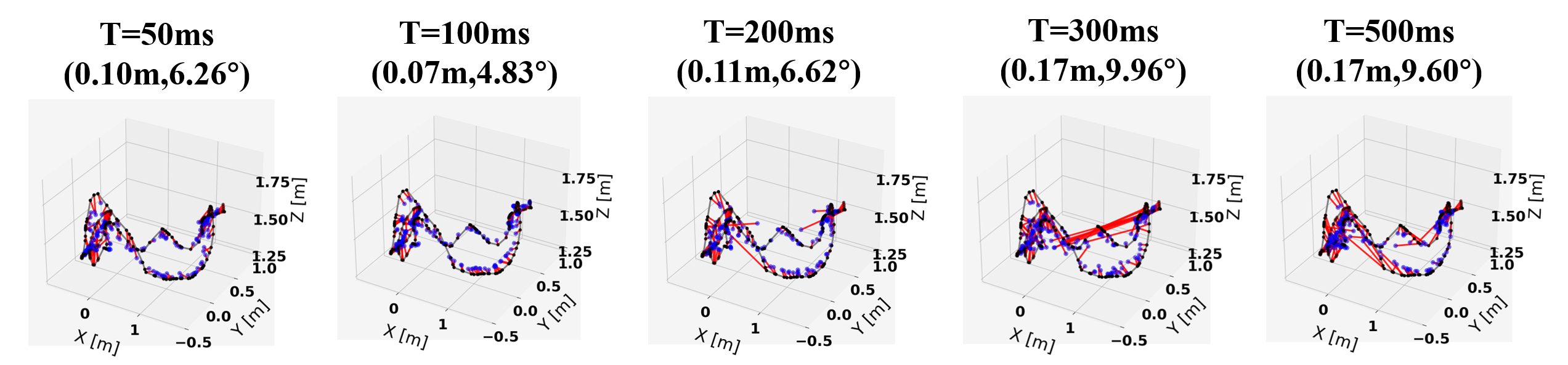}
\caption{Visualization of trajectories at different frame rates on Subject01. The ground truth trajectories are shown as black lines, while the decoded poses are represented by blue lines. The red dashed lines indicate the corresponding errors. There are average errors (translational error in meters, rotational error in degrees) below the time conditions.}
    \label{fig:exp2}
\end{figure}
As demonstrated in Fig.~\ref{fig:exp2-analysis} and Fig.~\ref{fig:exp2}, the accuracy of the decoded 6D pose from EEG signals is critically influenced by both the image presentation frequency and the visual response latency.  Notably, the results show that decoding accuracy peaks when images are presented at intervals of approximately 100 milliseconds. This observation appears to align with findings from Event-Related Potential (ERP) analysis, where the P1 wave—typically emerging around 100 ms after the onset of visual stimuli—serves as a key marker of early-stage visual processing in the brain.

Interestingly, this temporal resonance is not only reflected in decoding performance but also aligns with subjective reports from participants. Many noted that playback at 100 ms intervals felt “immersive” and neither too fast nor too slow, indicating a natural perceptual comfort zone.

When the timing of visual stimulus presentation closely matches the latency window of the P1 wave, decoding performance reaches its optimal level. This temporal alignment suggests that the brain forms spatial representations of visual input in roughly 100 ms cycles. Consequently, it provides indirect evidence that the spatial cognition system may operate on a similar timescale. The synchronization between stimulus timing and the brain's natural processing rhythm not only enhances decoding precision but also supports the notion that spatial perception from visual cues may be organized in discrete temporal units centered around the 100 ms mark.

\subsection{How Are Spatial EEG Patterns Across Electrodes Involved in 6D Pose Decoding?}

To explore the spatial organization of neural signals underlying 6D pose decoding, we analyzed the relationship between EEG electrode locations and decoding performance using two complementary approaches: (1) gradient-based attribution maps to identify channel-wise contributions, and (2) scalp energy topographies to characterize spatiotemporal dynamics of EEG activity.

\subsubsection{Attribution Map of EEG Topography}
\begin{figure}[t]
    \centering

    \begin{subfigure}{0.482\linewidth}
        \centering
        \includegraphics[width=\linewidth]{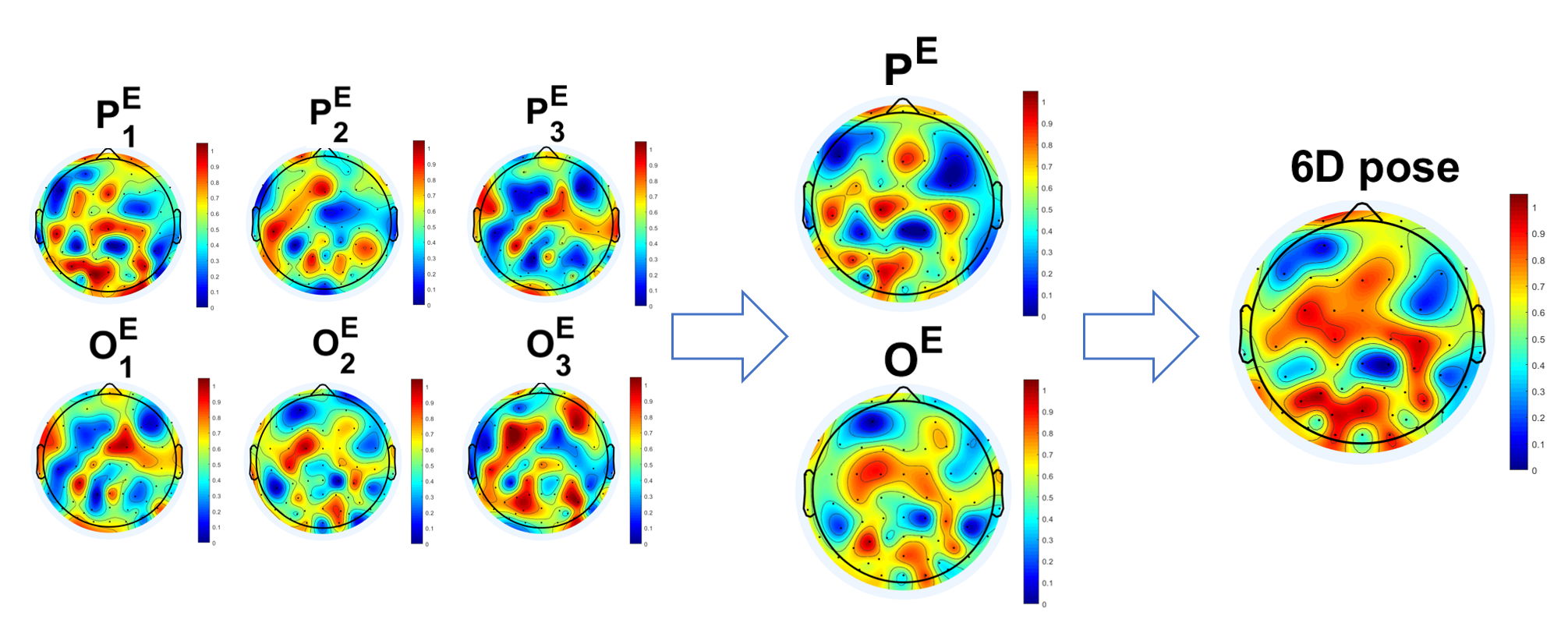}
        \caption{All subjects}
        \label{fig:all_subjects}
    \end{subfigure}
    \hfill
    \begin{subfigure}{0.48\linewidth}
        \centering
        \includegraphics[width=\linewidth]{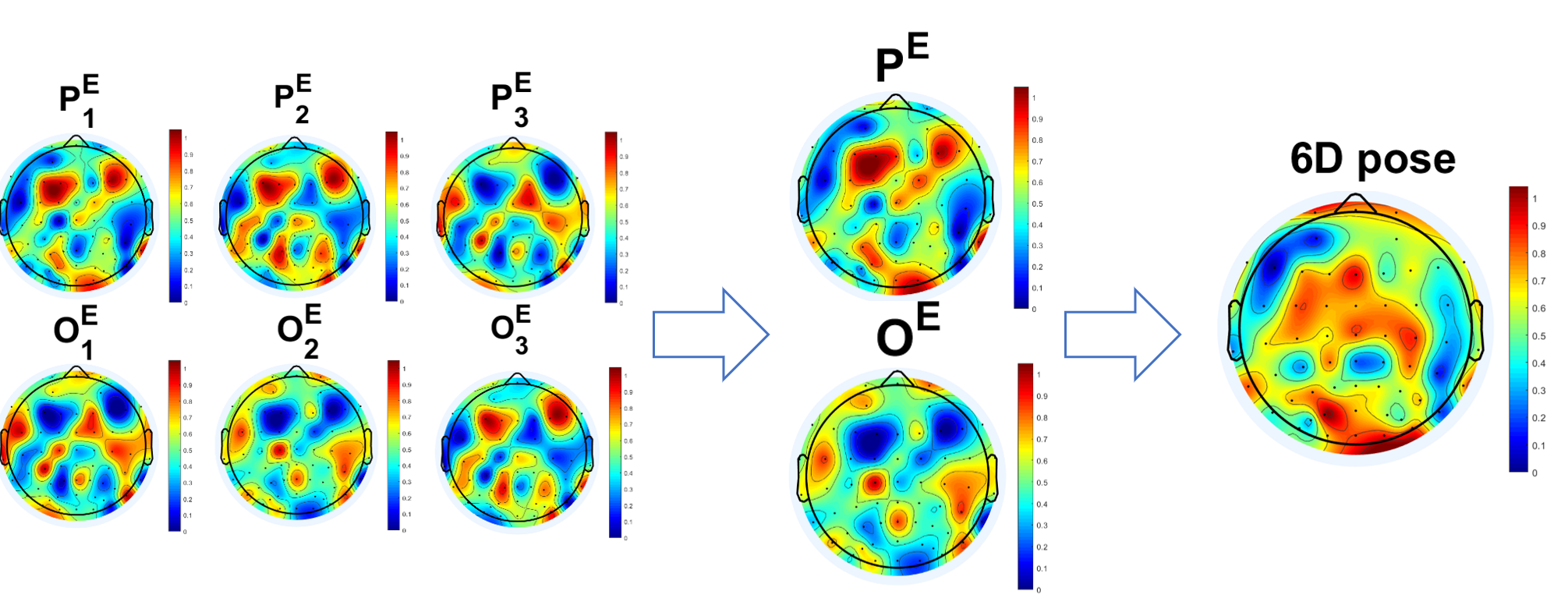}
        \caption{Subject 1}
        \label{fig:sub1}
    \end{subfigure}
    \vspace{0.3cm}

    \begin{subfigure}{0.48\linewidth}
        \centering
        \includegraphics[width=\linewidth]{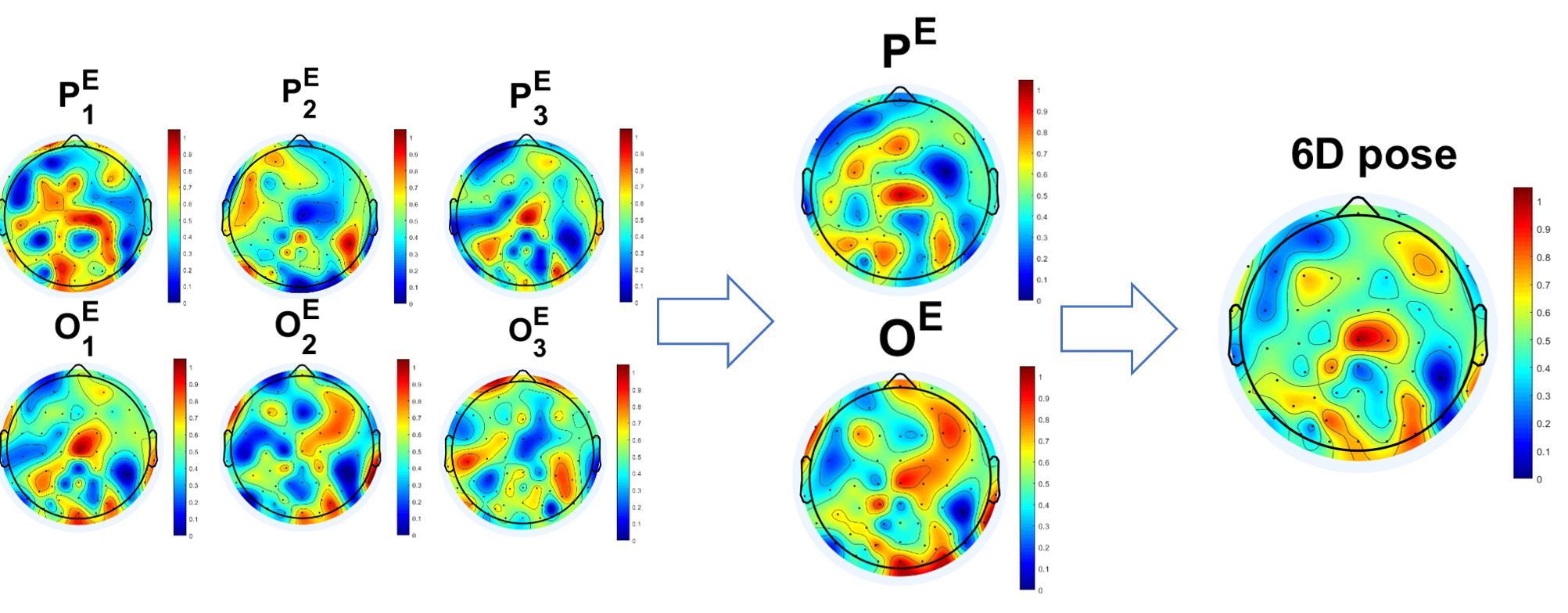}
        \caption{Subject 2}
        \label{fig:sub2}
    \end{subfigure}
    \hfill
    \begin{subfigure}{0.48\linewidth}
        \centering
        \includegraphics[width=\linewidth]{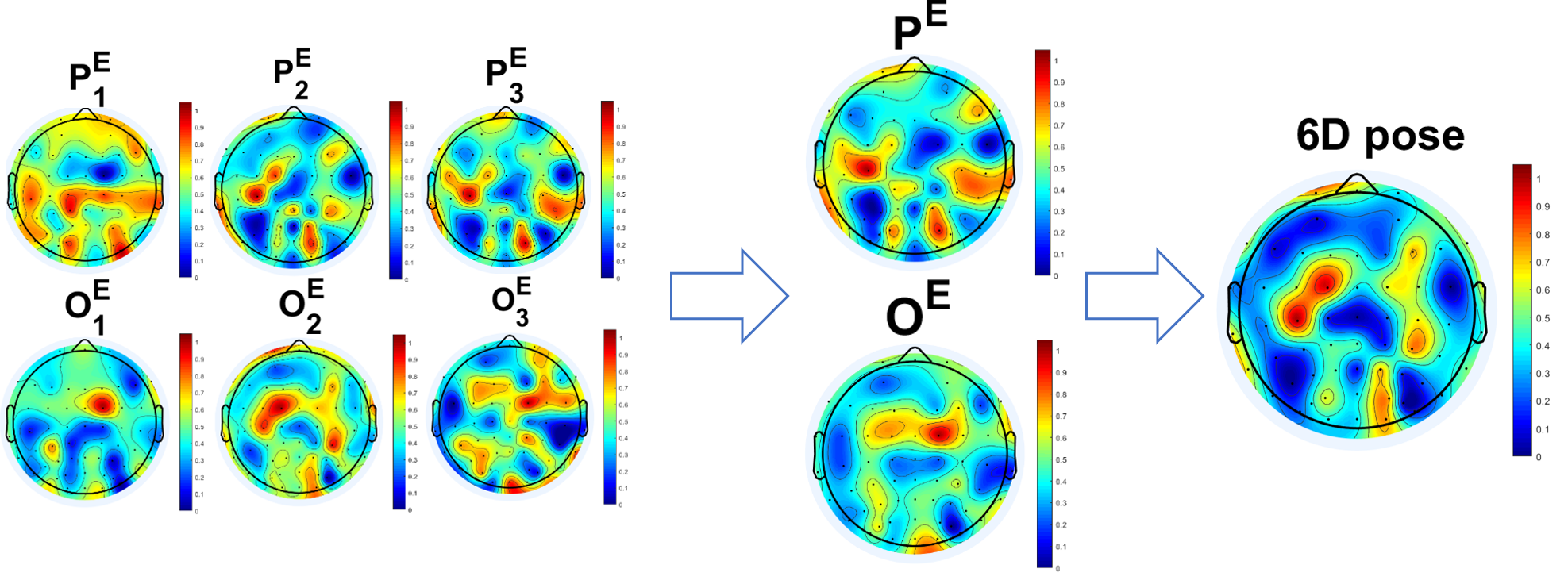}
        \caption{Subject 3}
        \label{fig:sub3}
    \end{subfigure}
    \vspace{0.3cm}

    \begin{subfigure}{0.48\linewidth}
        \centering
        \includegraphics[width=\linewidth]{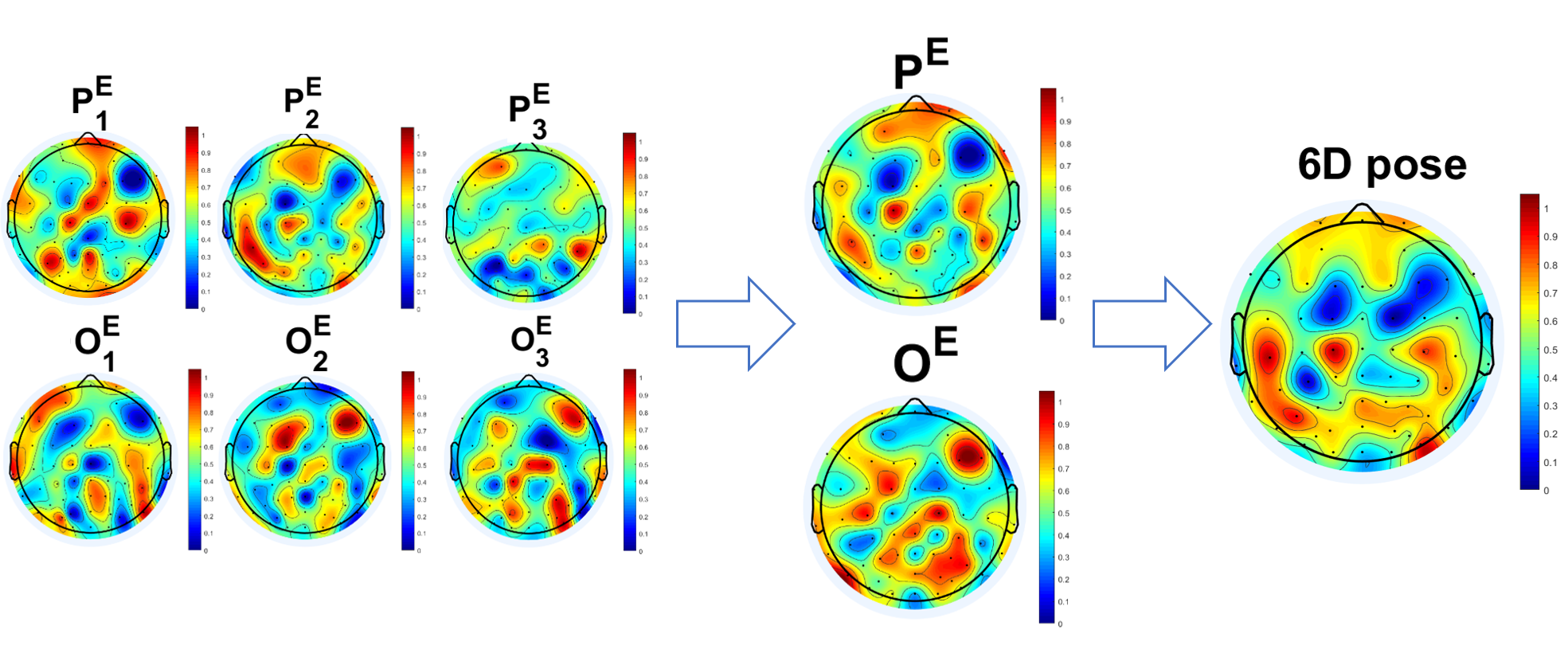}
        \caption{Subject 4}
        \label{fig:sub4}
    \end{subfigure}
    \hfill
    \begin{subfigure}{0.48\linewidth}
        \centering
        \includegraphics[width=\linewidth]{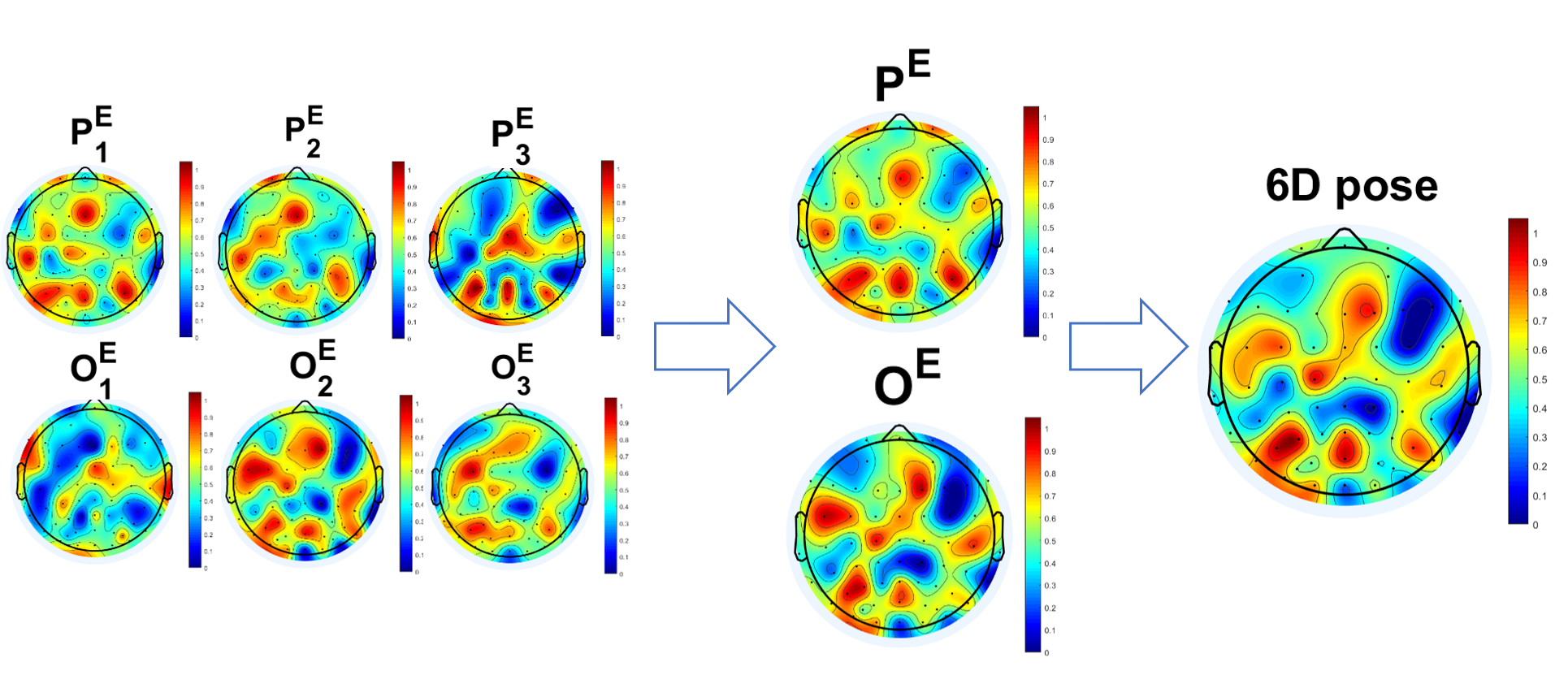}
        \caption{Subject 5}
        \label{fig:sub5}
    \end{subfigure}

    \caption{6D pose-related EEG topography attribution map. $\mathbf{p}^E_1$ denotes the first value of $\mathbf{p}^E$, which is analogous to $\mathbf{o}$. (a) across all subjects. (b–f) Individual results from five participants. Position and orientation-related components exhibit distinct spatial distributions, with consistent patterns across subjects and complementary features among orientation axes.}
    \label{fig:eeg_all_and_individuals}
\end{figure}

To identify the EEG channels most relevant for decoding, we performed attribution analysis by computing the gradient of the model output with respect to the input EEG signals. For each batch, gradients were enabled on the input tensor, and the model was forward-propagated to obtain predictions. A loss was defined with respect to a target dimension, and backpropagation yielded the gradient of the loss with respect to the input, capturing the contribution of each EEG channel to the prediction.

As shown in Fig.~\ref{fig:eeg_all_and_individuals}, the spatial distributions of neural relevance differ markedly between position and orientation decoding. Position-related signals are primarily concentrated near central electrodes, suggesting engagement of midline sensorimotor or parietal regions. In contrast, orientation decoding relies more heavily on lateral electrodes, indicating lateralized processing. 

Interestingly, one of the orientation components shows a complementary activation pattern relative to the other two, indicating possible neural orthogonality in encoding distinct rotational axes. Moreover, three out of five participants demonstrate clear spatial dissociation between electrodes relevant for position versus orientation decoding, while the remaining two show a more convergent pattern. These individual differences highlight the interplay between universal and idiosyncratic neural strategies for spatial representation.

Together, these findings demonstrate that gradient-based attribution analysis provides a data-driven approach to spatially distinct neural substrates underlying different components of spatial recognition representation.

\subsubsection{EEG Scalp Energy Topography}
\begin{figure}
\centering
\includegraphics[width=0.75\linewidth]{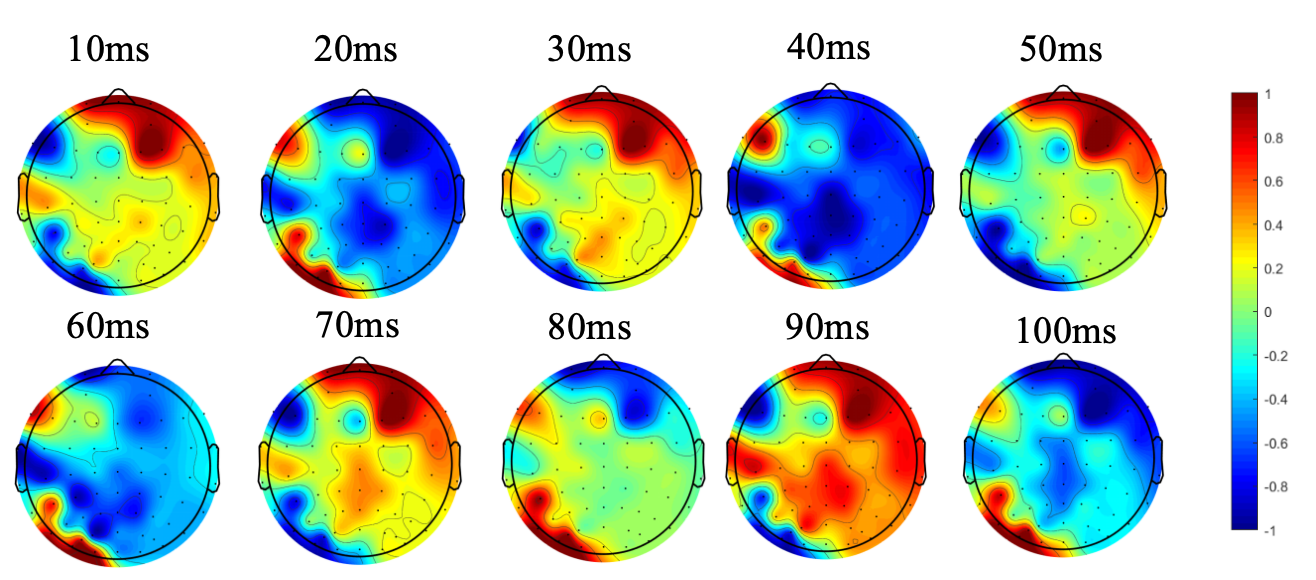}
\caption{EEG scalp topographies at 10 ms intervals from 10 to 100 ms following the onset of a spatial sequence frame. The maps reveal a structured cascade of activation and a posterior rhythmic pattern consistent with low-frequency perceptual sampling dynamics.}
\label{fig:spatial}
\end{figure}

In this process, the spatial domain analysis of EEG signals aims to display the brain's electrical activity changes within a 100ms time window while viewing an image. To capture these changes, the time window is divided into 10 intervals. Within a single 100ms frame of a spatial sequence, the EEG topographies reveal a reproducible spatiotemporal progression of cortical activation, shown in Fig. \ref{fig:spatial}. An initial lateralized occipital response rapidly transitions into widespread parietal negativity, followed by a rebound of central positivity. This sequence suggests temporally phased recruitment of sensory and higher-order areas during early perceptual parsing of spatial inputs.

Superimposed on this progression is a low-frequency oscillatory dynamic, characterized by polarity reversals over posterior electrodes with an approximate periodicity of ~60 ms. These rhythmic fluctuations may reflect an intrinsic temporal sampling mechanism that supports the integration of dynamic spatial information. The phase-aligned transitions suggest coordinated activity across occipito-parietal networks engaged in sequential spatial updating.
\begin{figure}[t!]
\centering
\includegraphics[width=0.8\linewidth]{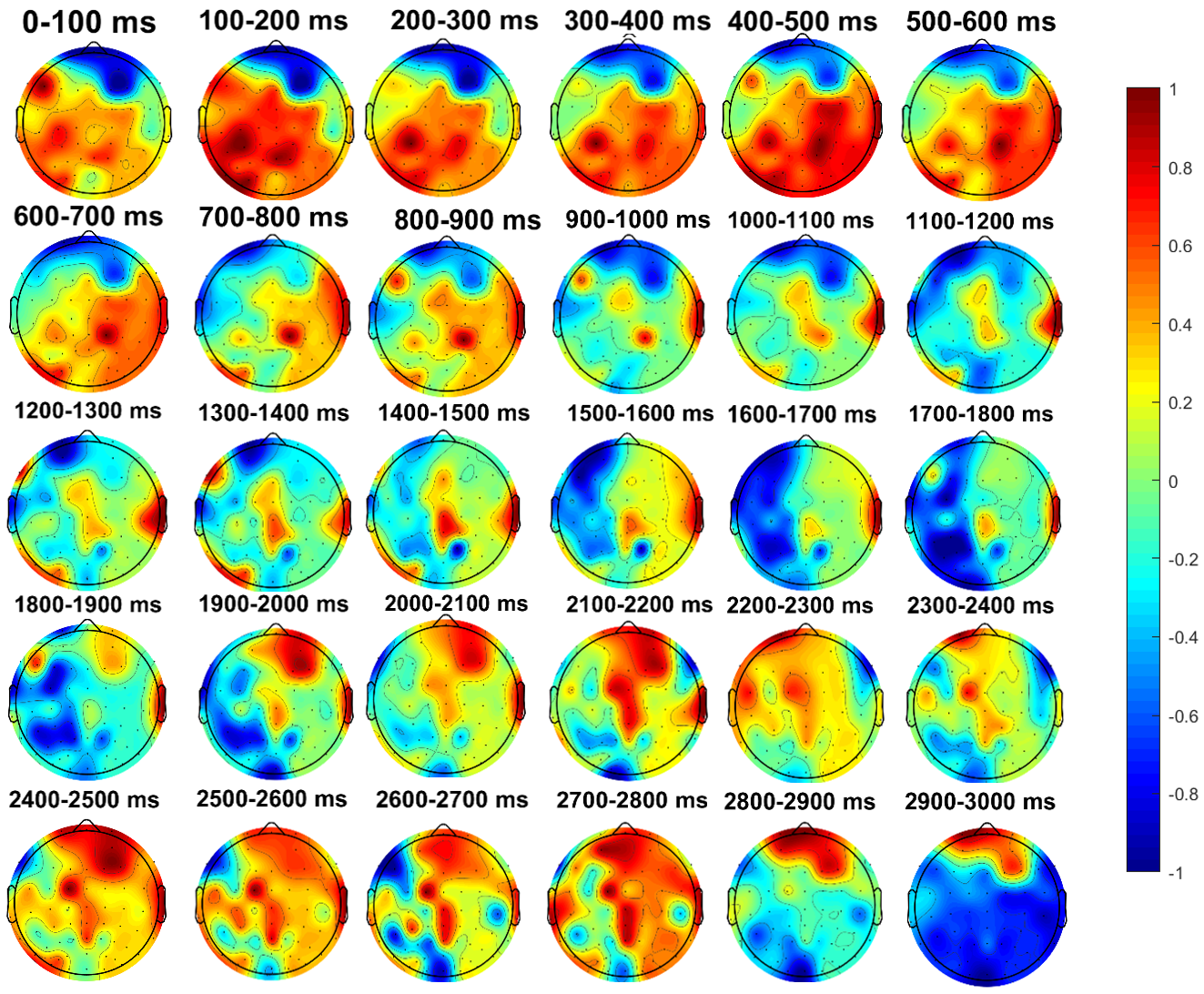}
\caption{EEG topographies illustrating spatiotemporal dynamics during spatial sequence viewing. Neural responses to 100-ms single-image segments reveal a time-locked posterior positivity followed by occipito-parietal polarity reversal, recurring rhythmically throughout the 3-s trial.}
\label{fig:spatial1a}
\end{figure}

\begin{figure}[t!]
\centering
\includegraphics[width=0.8\linewidth]{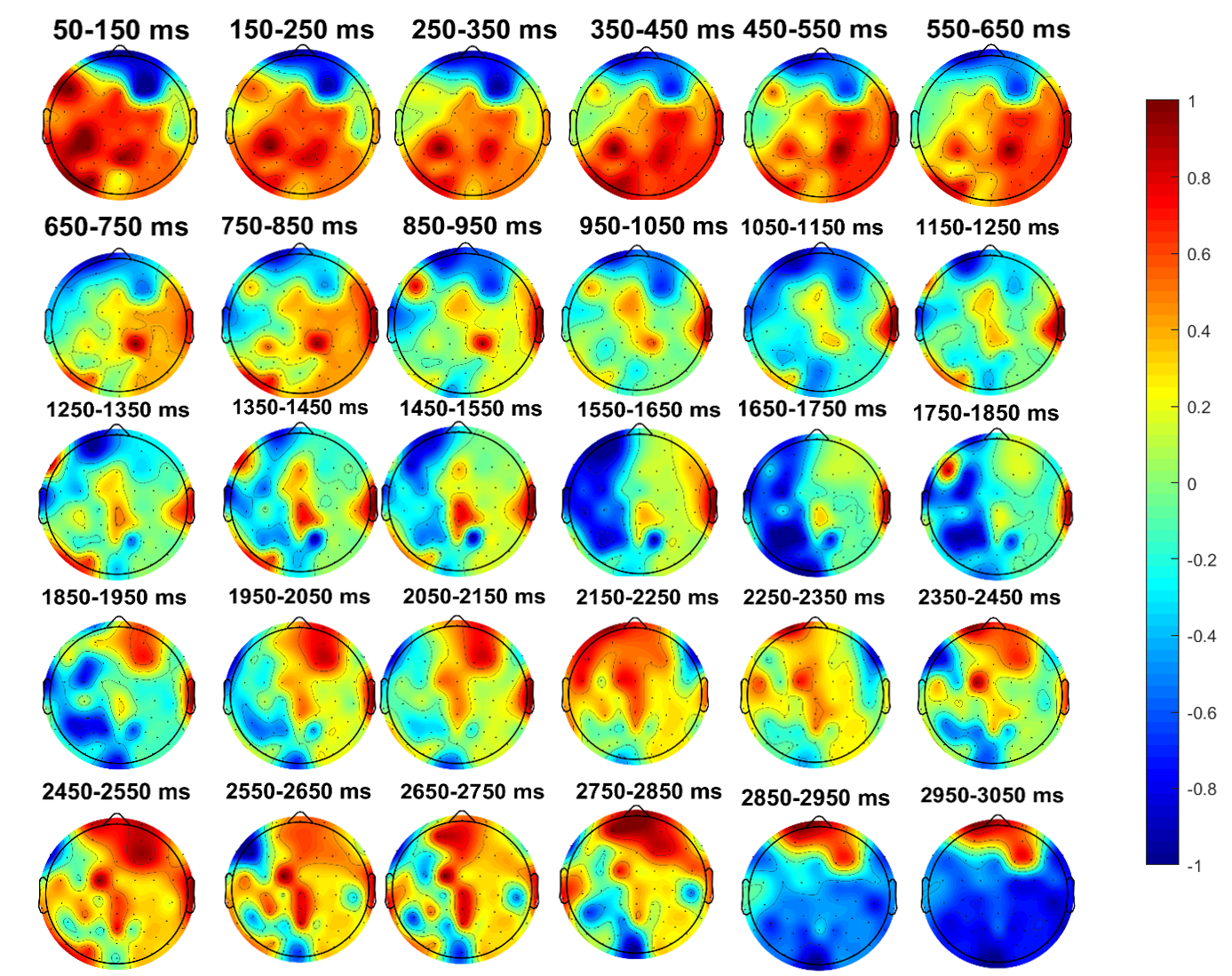}
\caption{EEG topographies during 100-ms transition segments across image boundaries (50 ms before and after transition). These evoke amplified parietal negativity and attenuated rebound activity, consistent with phase resetting during perceptual updating.}
\label{fig:spatial1b}
\end{figure}

In the second part of the experiment, we examined the brain’s time-domain dynamics during continuous spatial image sequence viewing, using a 100-ms analysis window. Two types of temporal segments were defined: the first segment corresponds to a full 100ms of an single image being played. The second segment spans across the transitions between images, where the 100ms segment includes 50ms from the previous image and 50ms from the following one. This segmentation strategy enabled us to dissociate the neural correlates of sustained visual processing from those associated with perceptual updating. As shown in Fig. \ref{fig:spatial1a} and Fig. \ref{fig:spatial1b}, and consistent with the fast-scale dynamics observed within the initial 100 ms, single-image responses (panel a) exhibit a time-locked cascade beginning with a posterior positivity and subsequent occipito-parietal polarity reversal peaking around 150–250 ms. This early pattern recurs rhythmically every ~600 ms throughout the 3s sequence, suggesting a sustained internal sampling cycle likely aligned with band oscillations. These periodic activations reflect the brain’s intrinsic mechanism for maintaining perceptual continuity in temporally discrete visual input. In contrast, transitions between images (panel b) evoke amplified and prolonged parietal negativity during early stages (~650–1250 ms), accompanied by a transient disruption of the endogenous rhythmic cycle. While the initial posterior response remains preserved, the oscillatory rebound is attenuated and delayed, indicating a phase resetting process engaged by image transitions. These findings suggest that spatial sequence perception relies on an internally structured, rhythmically governed temporal parsing mechanism that remains stable under continuous input but flexibly resets to accommodate dynamic perceptual boundaries.

\section{Discussion}
In this study, we moved beyond traditional, synthetic visual stimuli and instead employed naturalistic egocentric video viewing to investigate spatial cognition. Through a series of carefully designed experiments, we demonstrated that non-invasive EEG signals contain rich information reflecting human spatial representations evoked by egocentric videos. Compared to invasive methods, the use of non-invasive brain-computer interfaces (BCIs) opens new avenues for studying spatial cognition by providing convenient access to macroscopic brain functional representations.

Our results further reveal that spatial representations emerges robustly only when the visual stimuli preserve spatial context in a continuous and coherent manner. This finding aligns with everyday experiences—such as the common disorientation reported during 3D gaming when spatial continuity is disrupted—highlighting the ecological validity of our approach.

Moreover, we introduced a novel method to decode fine-grained 6D pose information from scalp EEG signals, confirming that spatial cognition information is represented across distributed brain regions accessible via EEG. The improved decoding techniques applied here can effectively extract informative signal components from the inherently noisy EEG data, highlighting the potential of data-driven approaches to reveal latent neural patterns underlying spatial cognition.

Another key insight is that an approximately 100 ms video frame rate optimally supports the emergence of spatial representations in EEG, resulting in high-fidelity 6D pose decoding. This temporal window is consistent with participants’ subjective reports of immersive spatial experience and corresponds with known visual processing dynamics, such as the timing of the P1 ERP component.

Beyond decoding, we expanded the analytical framework for EEG signals by employing data-driven methods, including attribution maps, to uncover neural mechanisms underlying spatial cognition. This approach transcends traditional waveform or frequency inspection, offering objective insights into spatially distinct neural substrates and demonstrating the power of neural networks in interpreting complex EEG data.

Beyond advancing our understanding of spatial cognition in EEG, this research has important implications for robotics and autonomous systems. Spatial cognitive capability is equally critical for robots, forming the foundation for autonomous systems and enabling a wide range of applications, including virtual reality~\cite{burdea2024virtual}, delivery drones~\cite{bamburry2015drones}, and autonomous driving~\cite{yurtsever2020survey}. Robust localization is essential for robots, enabling them to understand the spatial characteristics of their environment. Localization~\cite{lowry2015visual,garg2021your,mur2017orb,engel2017direct,kendall2016modelling,wang2020atloc} is a fundamental task in robotics, with visual SLAM serving as a prime example, having evolved over several decades~\cite{cadena2016past}. Despite significant advancements driven by deep learning~\cite{teed2021droid,teed2024deep}, which can yield precise localization results, its robustness in complex environments still falls short of human capabilities~\cite{cadena2016past,dai2020rgb}. As a result, some researchers have turned to brain-inspired mechanisms~\cite{shen2023orb}, exemplified by models like RatSLAM~\cite{milford2004ratslam} and NeuroSLAM~\cite{yu2019neuroslam}, which incorporate various types of neural cells to achieve preliminary yet robust localization. Efforts to integrate brain-like chips are also underway to further advance this field~\cite{yu2023brain}. However, due to the nascent state of neuroscience research~\cite{georges1999spatial,rolls2005spatial,bats2008representation,finkelstein2015three,ginosar2021locally,omer2023contextual,alexander2023rethinking}, these efforts remain in the early stages. Given that our approach extracts spatial cognitive features directly from EEG signals in a data-driven manner, our findings may significantly advance the development of brain-inspired navigation systems.

This study also has limitations. Our paradigm involved passive 2D video viewing, which lacks the depth and multisensory integration of naturalistic 3D environments. Subsequent studies will aim to incorporate virtual reality paradigms with synchronized vestibular, proprioceptive, and auditory cues to better mimic real-world spatial perception and enhance the quality of spatial cognition signals in EEG. 
In future work, we also aim to extend our paradigm by incorporating eye-tracking data to investigate brain-eye coordination during dynamic spatial perception. This will allow us to examine how oculomotor behavior interacts with neural dynamics to support real-time spatial cognition in ecologically valid settings.
\section{Method}

\begin{figure}[t!]
    \centering
    \includegraphics[width=\linewidth]{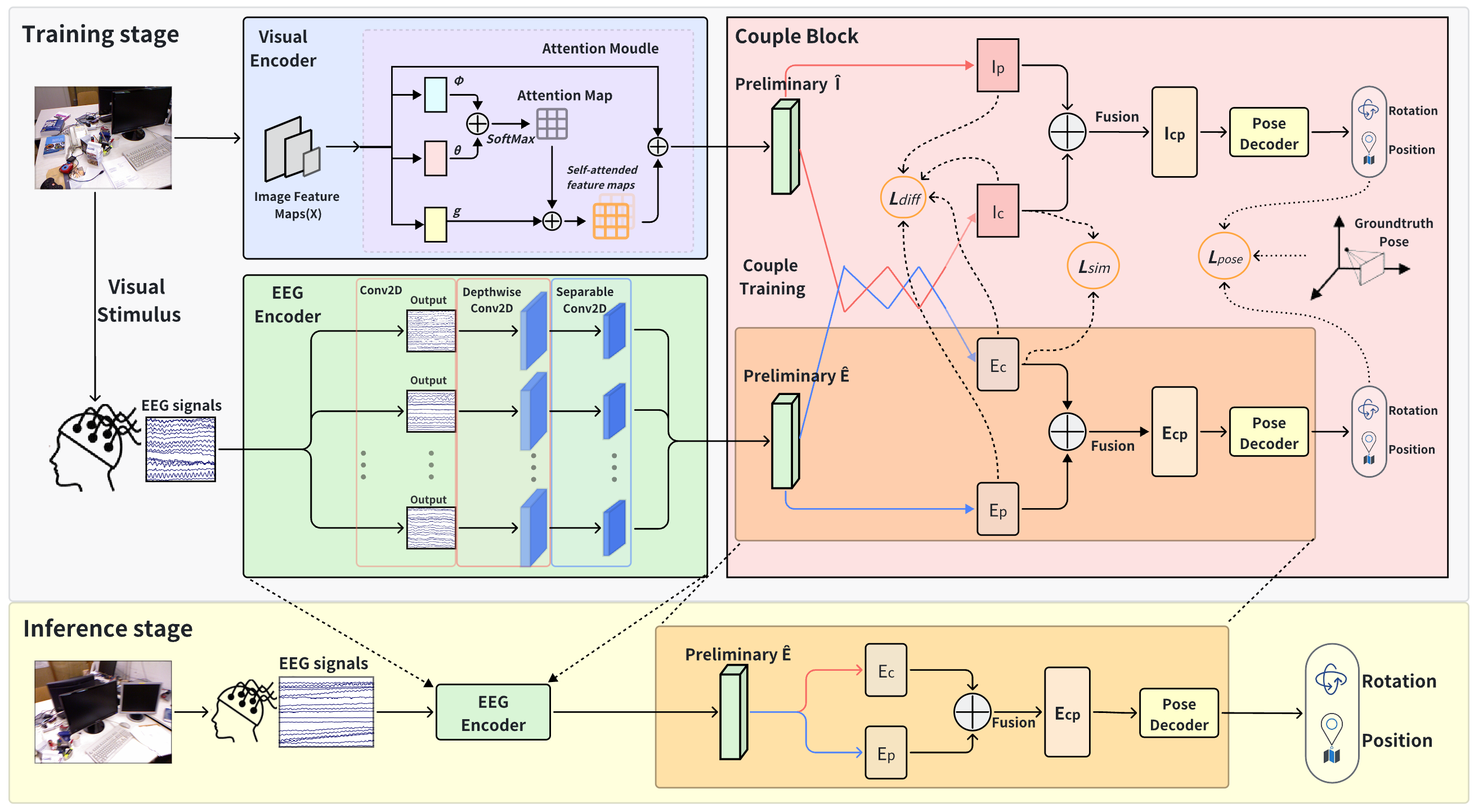}
    \caption{Overview of Decoding 6D pose from EEG signals using coupled training with visual
guidance. The preliminary representations of EEG and visual are extracted by their encoder respectively. Representations undergo coupled training through couple block. After training, using the inference stage for testing.}
    \label{fig:overview}
\end{figure}

This section describes the process of decoding a 6D pose from EEG signals. When subjects observe scene image sequences stimuli, EEG signals can capture brain activity related to spatial cognition. The proposed method decodes these EEG signals to infer the 6D pose as recognized by the subject. There are training and inference stages in the proposed method. To address the challenge of the low signal-to-noise ratio in EEG signals, the EEG model is coupled trained with a visual model to enhance the ability of 6D pose decoding during training stage. As shown in Fig.~\ref{fig:overview}, in the proposed method, preliminary feature extraction is performed on each EEG signal segment and its corresponding observed image using an EEG encoder and a visual encoder, respectively. These EEG features are then guided by the visual features to align with the 6D pose decoding task. After training, in the inference stage, the EEG model can generate high-quality features solely based on the EEG signals, which are used by a pose decoder to predict the pose, focusing on both 3D position and 3D orientation.

\subsection{EEG Data Acquisition}
\label{sec:eeg_acquisition}

\begin{figure}[t!]
    \centering
    \includegraphics[width=\linewidth]{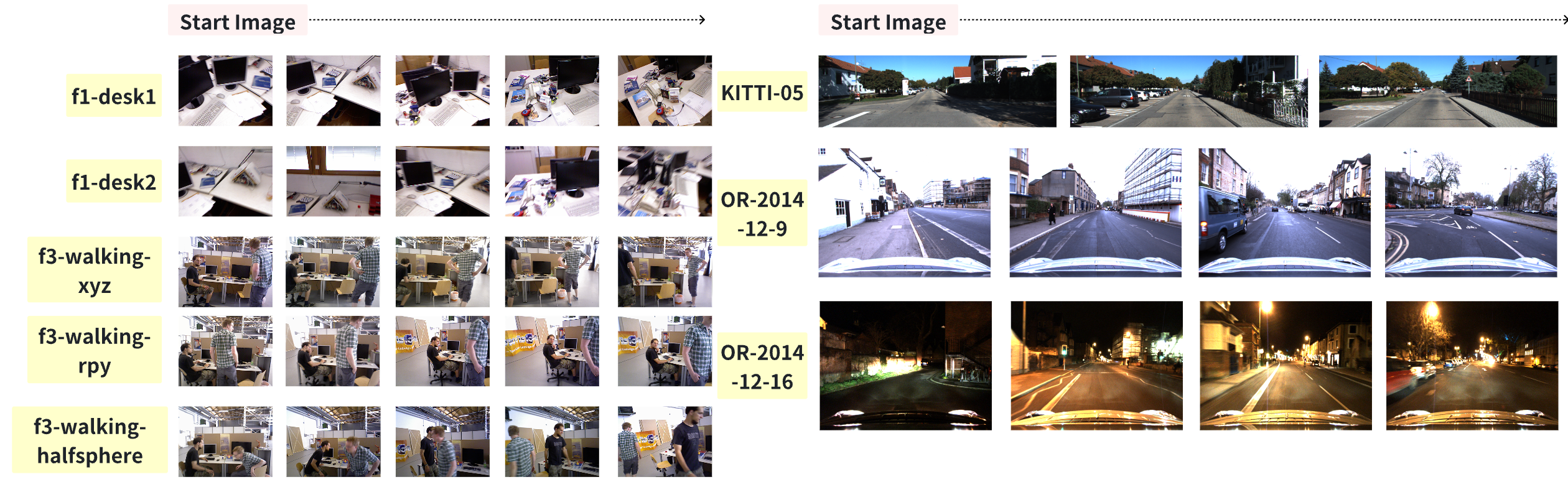}
\caption{Visualization of images from different scenes in the dataset.} 
    \label{fig:dataset1}
\end{figure}

Datasets pairing EEG with image sequences containing spatial information are quite limited. Therefore, we constructed a new dataset. 
We conducted data collection under various conditions, including sequential and random image presentations, different subjects, and multiple egocentric videos captured from diverse scenes. 
The egocentric video viewed by participants consisted of indoor datasets from the TUM RGB-D benchmark—including f1-desk1, f1-desk2, f3-walking-xyz, f3-walking-rpy, and f3-walking-halfsphere \cite{sturm2012benchmark}—as well as outdoor datasets from the KITTI-05 \cite{geiger2013vision} and Oxford RobotCar (2014-12-09 and 2014-12-16) benchmarks\cite{maddern20171}. 
The corresponding video data layouts are illustrated in Fig.~\ref{fig:dataset1}.
Subjects are instructed to focus their attention, view the sequential images, and engage in mental imagery of navigating through the space, with each camera image displayed at experimentally controlled frequencies. During data collection, timestamps for both the image presentation and EEG signal acquisition are recorded to achieve precise temporal alignment. This alignment enables the extraction of EEG segments corresponding to each image stimulus, thereby completing the pairing of EEG data with images.

\begin{figure}[t!]
     \centering
      \includegraphics[width=0.5\linewidth]{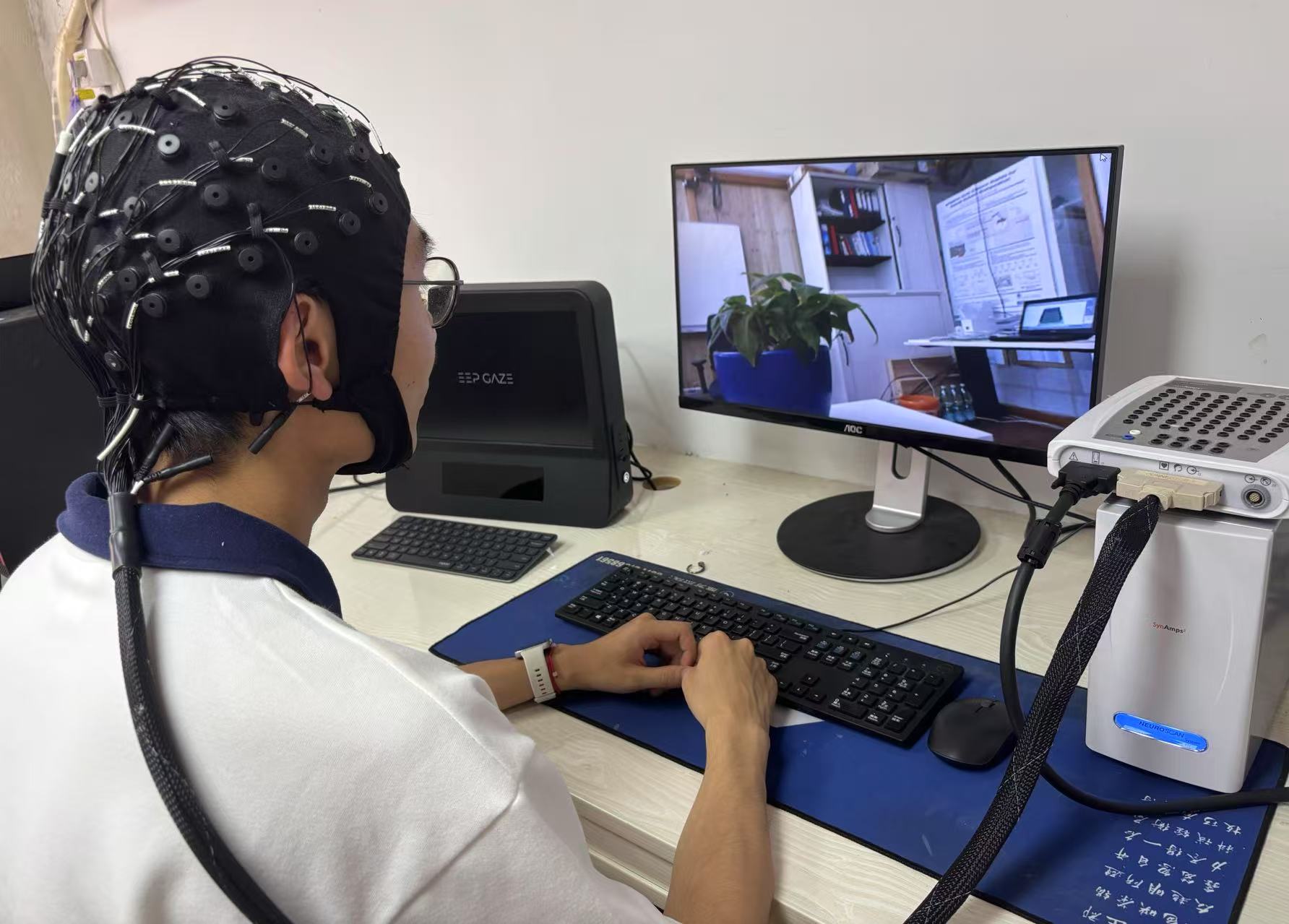}
       \caption{Experiment configuration.}  
       \label{fig:Electrode-placement}
 \end{figure}

The EEG signals were primarily acquired using a 64-channel NeuroScan EEG cap, which includes 60 surface electrodes that cover the entire scalp along with reference, ground, and additional functional channels. Each electrode is positioned according to the international 10-20 standard, with a sampling rate set at 1000Hz. The subjects sat in a comfortable chair at an appropriate distance and viewing angle from the screen, with the screen brightness and contrast adjusted to suit each individual preferences. These settings were tailored for each subject to ensure they could view the image sequences in a comfortable and familiar manner. The experiment configuration is provided in Fig.~\ref{fig:Electrode-placement}.

After completing the EEG signal acquisition, signal preprocessing is conducted to enhance the signal-to-noise ratio of the EEG signal. EEG data were first band-pass filtered with cut-off frequencies of 1Hz and 75Hz. Following this, the common average reference (CAR) method is applied, where the mean signal across all electrodes was subtracted from the signal of each individual electrode. Independent Component Analysis (ICA) is then performed to extract independent components, using the EEGLAB toolbox~\cite{delorme2004eeglab}. Artifact-related components are identified using ICLabel plugin within EEGLAB, combined with visual inspection. These artifact components are removed, and the remaining components are used to reconstruct the EEG signals. The paradigm parameters are summarized in Table~\ref{table:paradiam_parameters}.

\begin{table}[t]  
\centering  
\caption{Experimental Paradigm Parameters}  
\begin{tabular}{c|c}  
\toprule  
\textbf{Parameter} & \textbf{Value} \\ \midrule  
EEG Sampling rate & 1000 hz\\ 
Number of electrodes & 60\\ 
Filtered frequency & 1-75 hz\\ \bottomrule  
\end{tabular}
\label{table:paradiam_parameters}
\end{table}

\subsection{Training Stage}

\subsubsection{Network}
Based on the duration of each image stimulus, the collected EEG signals are segmented to obtain EEG segments corresponding to each visual stimulus image. These paired data are then used as inputs during the training phase. Each modality input is processed by  EEG and visual encoders, respectively. The EEG encoder, composed of multiple Conv layers, processes the input EEG data to extract initial EEG features $\bar{\mathbf{E}}$. To extract preliminary image features, we adopt residual networks along with an Attention module as the core of the visual encoder. The Attention module can intelligently assign importance weights to various features, enabling the model to focus on the most critical features for pose regression. Ultimately, the output $\bar{\mathbf{I}}$ of the visual encoder is generated from a 4096-dimensional fully connected layer.

Based on $\bar{\mathbf{E}}$ and $\bar{\mathbf{I}}$, a coupled FC (a single fully connected) and private FC networks respectively extract coupled and private features from the initial features in each modality, as follows:
\begin{equation}
    \begin{aligned}
        \mathbf{E}_c &= FC_{(I,E)}(\bar{\mathbf{E}}) \\
        \mathbf{I}_c &= FC_{(I,E)}(\bar{\mathbf{I}}), 
    \end{aligned}
\end{equation}
where the same subscript ${(I,E)}$ indicates the FC network shares the same parameters across both modalities.
The private features for each modality, using their respective FC networks, are defined as:
\begin{equation}
    \begin{aligned}
        \mathbf{E}_p &= FC_{E}(\bar{\mathbf{E}})\\
        \mathbf{I}_p &= FC_{I}(\bar{\mathbf{I}}), 
    \end{aligned}
\end{equation}
where the subscripts ${I}$ and ${E}$ denote independent FC networks for each modality.

After obtaining the representations for coupled and private features, the features of the two types within the same modality are simply concatenated for subsequent pose regression. The fused representation is defined as :
\begin{equation}
\begin{aligned}
    \mathbf{E}_{cp}=[\mathbf{E}_c \oplus \mathbf{E}_p] \\
    \mathbf{I}_{cp}=[\mathbf{I}_c \oplus \mathbf{I}_p] \\
\end{aligned}
\end{equation}
Finally, four fully connected networks are used to predict the position and orientation in the pose for each modality.
\begin{equation}
    \begin{aligned}
        \mathbf{p}_{E} &= FC_{(P,E)}(\mathbf{E}_{cp}), \\
        \mathbf{o}_{E} &= FC_{(O,E)}(\mathbf{E}_{cp}), \\
        \mathbf{p}_{I} &= FC_{(P,I)}(\mathbf{I}_{cp}), \\
        \mathbf{o}_{I} &= FC_{(O,I)}(\mathbf{I}_{cp}), 
    \end{aligned}
\end{equation}
where $\mathbf{p}_{E}$, $\mathbf{o}_{E}$, $\mathbf{p}_{I}$, and $\mathbf{o}_{I}$ represent the position and orientation predictions from EEG signals and images, respectively. For the FC network, for example, the subscript $(P,E)$ in $FC_{(P,E)}$ denotes the FC network dedicated to estimating position from $\mathbf{E}_{cp}$ of the EEG signals.

\subsubsection{Training Loss}

There are three distinct types of losses. The first loss is associated with the output of the coupled FC  network, denoted as \(L_{sim}\). This loss is responsible for learning a coupled feature in the shared subspace between two domains. Using this loss, the cross-modal heterogeneity gap is minimized. The second type of loss, \(L_{diff}\), is designed for private features. These private features are learned in two private subspaces with distribution difference constraints, which help maximize the cross-modal heterogeneity gap. The third type of loss, \(L_{pose}\), is related to the pose
regression.
Based on the descriptions above, the overall learning of the model is accomplished by minimizing:
\begin{equation}
    Loss=L_{pose}+\alpha L_{sim}+\beta L_{diff}
\end{equation}
where \(\alpha\) and \(\beta\) are interaction weights that determine the contribution of each regularization component to the overall loss. Each of these component losses is responsible for achieving the desired subspace properties.

1) Coupled loss: To align two coupled features, we use TripletMarginLoss. In the coupled subspace, this loss function can reduce the difference between the coupled features of the two modalities, achieving optimal alignment. The loss for the coupled channel between the coupled representations of the two modalities is given by:
\begin{equation}
\begin{aligned}
    L_{\text{sim}} = \text{TripletMarginLoss}(\mathbf{E}_{c}^{a}, \mathbf{I}_{c}^{p}, \mathbf{I}_{c}^{n}) \\
    + \text{TripletMarginLoss}(\mathbf{I}_{c}^{a}, \mathbf{E}_{c}^{p}, \mathbf{E}_{c}^{n})
\end{aligned}
\end{equation}
Where $a$ is the anchor term, $p$ denotes the positive term of the same label (paired with the anchor), and $n$ denotes the negative term of a different label (unpaired with the anchor).

2) Private loss: The private loss encourages the encoding functions to extract specific information for each modality. The loss is defined through a soft subspace orthogonality constraint between the private features and the coupled features of each modality. In a training batch with multiple samples, let \(\mathbf{I}^m_c\) and {\(\mathbf{I}^m_p\) } and \(\mathbf{E}^m_c\) and \textit{\(\mathbf{E}^m_p\)} are $m$-th sample respectively in the two modalities.
The difference loss is calculated as:
\begin{equation}
L_{\text{diff}} =  
\left\| ({\mathbf{I}^m_c}^T \mathbf{I}^m_p) \right\|_F^2 +  
\left\| ({\mathbf{E}^m_c}^T \mathbf{E}^m_p) \right\|_F^2 +  
\left\| ({\mathbf{I}^m_p}^T \mathbf{E}^m_p) \right\|_F^2  
\end{equation} 
where  \(\left\| \cdot \right\|^2_F\)  denotes the square of the Frobenius norm. Besides the constraint between the coupled and private representations, a soft subspace orthogonality constraint between the private representations of the two domains is also added.

3) Task Loss: Each image has its corresponding ground-truth pose $[\mathbf{p}^{gt}, \mathbf{q}^{gt}]$, where $\mathbf{p}^{gt}$ represents the camera position and $\mathbf{q}^{gt}$ is the unit quaternion used to accurately describe orientation. Following \cite{brahmbhatt2018geometry}, the pose loss function for both modalities are following:
\begin{equation}
\begin{aligned}
    L_{pose}=\left\|\mathbf{p}^{gt}-\mathbf{p}\right\|_1 e^{-\delta}+\delta+\left\|\log \mathbf{q}^{gt}-\mathbf{o} \right\|_1 e^{-\gamma}+\gamma 
\end{aligned}
\label{equ:poseloss}
\end{equation}
where \(\delta\)  and \(\gamma\) are learnable weights used to balance the position loss and rotation loss. \(p'\)and \(q'\) represent the predicted position and unit quaternion, respectively. Since quaternions are not unique, logarithmic form of an unit quaternion q is defined 
\begin{equation}
\log \mathbf{q} =   
\begin{cases}   
 \frac{\mathbf{v}}{\left\|\mathbf{v}\right\|} \cos^{-1}{u} & \text{if }  \left\|\mathbf{v}\right\| \neq 0\\  
 0 & \text{otherwise}  
\end{cases}  
\end{equation}
where the unit quaternion q is composed of a scalar $u$ and a three-dimensional vector $\mathbf{v}$,  $q = (u, \mathbf{v})$.

\subsection{Inference Stage}

In the inference stage, the trained EEG model estimates the pose associated with the given EEG signals.
\begin{equation}
\begin{aligned}
    \mathbf{E}_{cp} & =[FC_{(I,E)}(\bar{\mathbf{E}})  \oplus FC_{E}(\bar{\mathbf{E}})] \\
    \mathbf{p}_{E} &= FC_{(P,E)}(\mathbf{E}_{cp}) \\
    \mathbf{o}_{E} &= FC_{(O,E)}(\mathbf{E}_{cp}),
\end{aligned}
\end{equation}
where $FC_{(I,E)}$ is fully coupled with the visual modality during the training stage, extracting 6D pose decoding information.

\bmhead{\textbf{Data availability}}
The image data are obtained from public datasets. EEG data will not be released to respect participant privacy, it can be provided upon reasonable request.

\bmhead{\textbf{Code availability}}
The code of the analysis of this work can be found in GitHub:https://github.com/HDU-ASL/EEG-BPD

\bibliography{sn-bibliography}

\end{document}